\documentclass[prl,aps,superscriptaddress,twocolumn]{revtex4-1}
\usepackage{graphicx,times}
\usepackage{bm}
\usepackage{amsmath}
\usepackage{amsfonts,subfigure,dsfont}
\usepackage{amssymb}
\usepackage{epsfig}
\usepackage{color}
\usepackage{amssymb,bbm,amsthm}
\begin{document}

\newcommand{\abs}[1]{\left\vert#1\right\vert}
\newcommand{\set}[1]{\left\{#1\right\}}
\newcommand{\bra}[1]{\left\langle#1\right\vert}
\newcommand{\ket}[1]{\left\vert#1\right\rangle}
\newcommand\braket[2]{\left.\left\langle#1\right|#2\right\rangle}
\newcommand{\Tr}[1]{\text{Tr}\left\{#1\right\}}
\def\I {{\rm 1} \hspace{-1.1mm} {\rm I} \hspace{0.5mm}}
\def\Z {{\mathds Z}}
\newcommand{\rosso}[1]{\color[rgb]{0.6,0,0} #1}

\title{The role of coherence in the  non-equilibrium thermodynamics of quantum systems}

\author{G. Francica}
\affiliation{Dip. Fisica, Universit\`{a} della Calabria, 87036
Arcavacata di Rende (CS), Italy} \affiliation{INFN - Gruppo
Collegato di Cosenza}

\author{J. Goold}
\affiliation{ICTP, Trieste, Italy}

\author{F. Plastina}
\affiliation{Dip. Fisica, Universit\`{a} della Calabria, 87036
Arcavacata di Rende (CS), Italy} \affiliation{INFN - Gruppo
Collegato di Cosenza}

\date{\today}

\begin{abstract}
Exploiting the relative entropy of coherence, we isolate the
coherent contribution in the energetics of a driven
non-equilibrium quantum system. We prove that a division of the
irreversible work can be made into a coherent and incoherent part,
which provides an operational criterion for quantifying the
coherent contribution in a generic non-equilibrium transformation
on a closed quantum system. We then study such a contribution in
two physical models of a driven qubit and kicked rotor.  In
addition, we also show that coherence generation is connected to
the non-adiabaticity of a processes, for which it gives the
dominant contribution for slow-enough transformation. The amount
of generated coherence in the energy eigenbasis is equivalent to
the change in diagonal entropy, and here we show that it fulfills
a fluctuation theorem.
\end{abstract}

\bigskip
\pacs{ }

\maketitle {\bf Introduction-} The characterization of
irreversibility and entropy production has been a central issue of
statistical mechanics since the inception of the theory and is by
now at the heart of modern approaches to non-equilibrium
thermodynamics. For quantum systems \cite{qtermo}, fluctuation
theorems \cite{storici,rev,CampisiRev} have been experimentally
tested \cite{experiments,expe2} and employed to study
irreversibility and to quantify its emergence~\cite{defner}, even
in the dynamics of closed systems, driven by external agents that
perform work~\cite{work1,work2,work3} or extract
it~\cite{extraction,ergotropy,daemonergo}. Understanding how
irreversibility emerges from microscopic laws is not only a
fundamental issue
\cite{kieu,tasaki2nd,siefert2nd,giovannetti,brandao}, it is also a
highly pragmatic one. This is particular evident in the current
effort to develop thermal machines from small quantum
systems~\cite{appli1,appli2,appli3,appli4}.

The role played by quantum coherence in thermodynamic settings
have been investigated extensively in the past few years
\cite{scully,coherence} but to our knowledge this role lacks a
formal clarification. The aim of this letter is to provide
precisely this.

Specifically, we will take advantage of the recently introduced
measures %which have been rigorously certified
to quantify coherence \cite{plenio,coherRev} and employ the so
called relative entropy of coherence to derive quantitative
relations between coherence generation and irreversibility.
Specifically, we will separate two contributions to the
irreversible work produced by a finite time driving, one due to
the generation of coherence and a second one due incoherent
transitions. Remarkably, we show that the each of the
contributions fulfills a fluctuation theorem. Furthermore, we
introduce a measure for the non-adiabaticity where the very same
coherent contribution can be isolated again, and for which it
gives the leading term in slow processes.
\begin{figure}[b]
        \begin{center}
        \includegraphics[width=0.8\columnwidth]{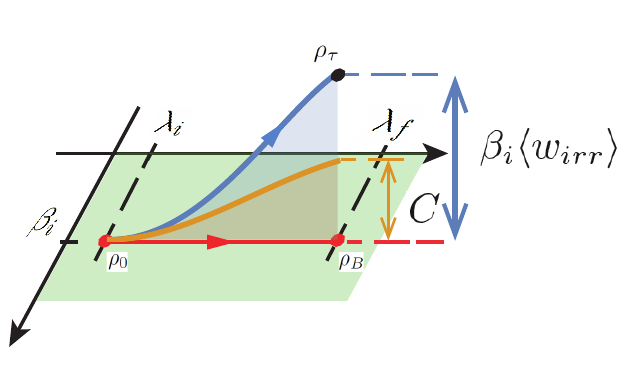}
        \includegraphics[width=0.8\columnwidth]{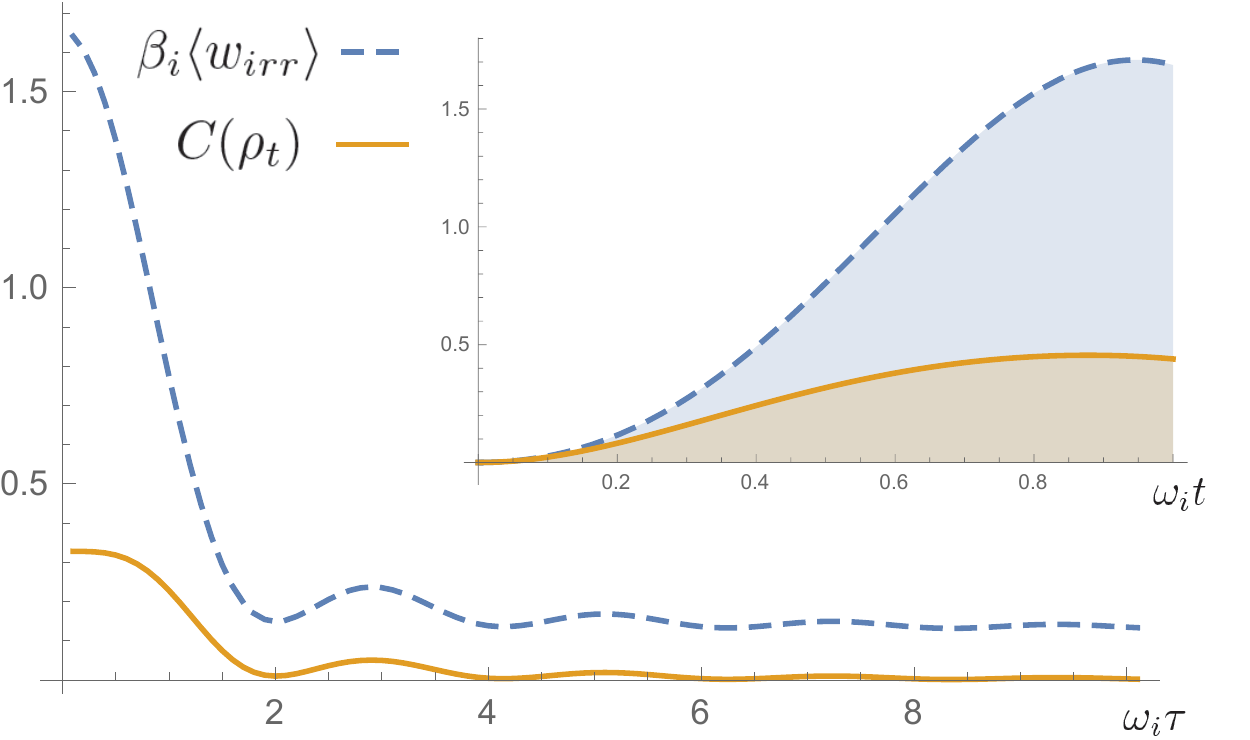}
\caption{(Color online). {\it Upper panel - } Sketch of the
process. The system starts at $t=0$ in the state $\rho_0$,
corresponding the red point $(\beta_i, \lambda_i)$ on the manifold
of equilibrium states (here represented as a plane). The red
horizontal line on the plane gives the set of instantaneous
equilibrium states at the same temperature. As time goes on,
$\lambda$ is changed from $\lambda_i$ to $\lambda_f$ and the
system goes out of equilibrium. The distance (measured with the
relative entropy) between $\rho_{\tau}$ and the equilibrium state
at $B=(\beta_i, \lambda_f)$ gives the irreversible work. A portion
of this distance (in orange in the plot) is due to the coherence
generated during the process. {\it Lower panel - } Irreversible
work and coherence produced for a spin undergoing the process
experimentally studied in Ref. \cite{expe2}, generated by the
Hamiltonian $H(t) = \hbar \omega(t) [\sigma^x \cos \varphi(t) +
\sigma^y \sin \varphi(t)]$, with $\varphi(t) = \pi t/2 \tau$, with
$\omega(t) = \omega_i ( 1 -t/\tau) + \omega_f t/\tau$ (we take,
here, $\omega_f=2 \omega_i$). The main plot shows both the
irreversible work and the coherence as a function of the process
duration $\tau$ (for $\beta_i \hbar \omega_i= 1$), while the inset
gives the instantaneous time dependence for $\omega_i \tau = 1,
\beta_i \hbar \omega_i = 2$.  }
        \label{scheme}
        \end{center}
\end{figure}
We will exclusively focus on generic unitary quantum process,
where systems are driven far from an initial equilibrium state
$\rho_0$ by an external driving described by some suitable
Hamiltonian. Here, the system is assumed to be closed, and the
time evolution operator $U_{\tau,0}[\lambda]$ is generated by the
Hamiltonian $H[\lambda(t)]$, which depends on a work parameter
$\lambda(t)$, changed from $\lambda(0)=\lambda_i$ to
$\lambda(\tau)=\lambda_f$ in the time interval $[0,\tau]$. The
instantaneous Hamiltonian can be written as $H[\lambda(t)]=\sum_n
\epsilon_{n}(t) \ket{n(t)}\bra{n(t)}$, and it changes from
$H_i\equiv H[\lambda_i]$ at $t=0$, into $H_f\equiv H[\lambda_f]$
at the final time $t=\tau$. Defining the family of free energies
$\beta F_{\beta, \lambda} = - \ln Z(\beta, \lambda)$, with the
partition function $Z(\beta, \lambda) = \Tr{e^{-\beta
H[\lambda]}}$, and the {\it instantaneous} equilibrium state,
$\rho^{\beta}_{\lambda} = e^{\beta (F_{\beta, \lambda} -
H[\lambda])}$, we can write the initial state as $\rho_0 =
\rho^{\beta_i}_{\lambda_i}$ and its time evolved (out of
equilibrium) counterpart as $\rho_\tau = U_{\tau,0}[\lambda]
\rho_0 U_{\tau,0}^\dag[\lambda]$. As we are assuming no energy
dissipation during the process, the increase in the average energy
of the system can be interpreted as work done on it by the driving
agent,
\begin{eqnarray}\label{WorkAverage}
\langle w \rangle &=& \Tr{H_f \rho_\tau} - \Tr{H_i \rho_0}\\
 &=& \sum_n \left( \rho_{nn}(\tau) \epsilon_{n}(\tau) - \rho_{nn}(0) \epsilon_{n}(0) \right)
\end{eqnarray}
where, from now on, every time we write a density matrix element, we understand it is evaluated in the instantaneous energy basis,
$\rho_{nm}(t) = \bra{n(t)} \rho_t \ket{n(t)}$. The initial population is, then $\rho_{nn}(0)= \exp \{\beta_i(F_{\beta_i,\lambda_i}-\epsilon_{n}(0)) \}$, while $\rho_{nn}(\tau)$  can be
expressed as $\rho_{nn}(\tau) = \sum_m \rho_{mm}(0)P_{m\rightarrow n}(\tau)$ \cite{polko08}, in terms of the transition probability $P_{m\rightarrow n}(\tau) =
\abs{\bra{n(\tau)}U_{\tau,0}[\lambda]\ket{m(0)}}^2$.

{\bf Irreversible work and coherence} - The energetic deviation
between the actual unitary evolution $\rho_0 \rightarrow
\rho_\tau$ and the quasi-static isothermal one is known as
irreversible work. We would like to stress that the term can be
misleading, but in this context we assume that, by definition, the
irreversible work is energy which is not recoverable. An
isothermal transformation (unlike, e.g., the adiabatic one) is not
unitary and would bring the system through a path on the manifold
of equilibrium states up to a final state with the same initial
inverse temperature $\beta_i$, and final Hamiltonian, $\rho_B
\equiv \rho^{\beta_i}_{\lambda_f} = \exp \{\beta_i(F_{\beta_i,
\lambda_f} - H_f)\}$, see the sketch in Fig.~\ref{scheme}. The
work performed in this isothermal process is given by the free
energy change, $\Delta F = F_{\beta_i, \lambda_f} - F_{\beta_i,
\lambda_i} \equiv F_B - F_i$, and comparing the actual and ideal
works, we obtain the so called irreversible work (and entropy),
\begin{equation*}
\langle S_{irr} \rangle = \beta_i \langle w_{irr} \rangle :=
\beta_i (\langle w \rangle - \Delta F) =   D(\rho_\tau || \rho_B)
\end{equation*}
where the last equality expresses the irreversible entropy as the
Kullback-Leibler divergence between the actual final state
$\rho_\tau$ and the final reference state $\rho_B$, $D(\rho_\tau
|| \rho_B)= - S(\rho_\tau) - \Tr{\rho_\tau \ln \rho_B}$,  with
$S(\rho)$ being the Von Neumann entropy $S(\rho)=-\Tr{\rho\ln
\rho}$.

The irreversible work can be decomposed into two contributions,
coming from i) unwanted transitions (or, more precisely, from the
difference in populations between $\rho_{\tau}$ and $\rho_B$), and
ii) the coherence generated by the unitary driving during the
actual process ($\rho_B$, instead, is fully incoherent). Let us
first state our first result which we will then prove:
\begin{equation}
\langle S_{irr} \rangle =  C(\rho_\tau) +
D(\Delta_{\tau}[\rho_\tau] ||\rho_B)  \label{theo1}
\end{equation}
where, the amount of coherence denoted by $C$, is quantified by
the relative entropy of coherence of $\rho_{\tau}$ which is
defined as
\begin{equation}\label{CoherMeas}
 C(\rho_\tau) = \text{min}_{\sigma\in I_\tau} D(\rho_\tau||\sigma)
\end{equation}
where $I_\tau$ is the set of incoherent states in the final energy
basis $\{\ket{n(\tau)}\}$. The second term in Eq.~(\ref{theo1}),
instead, quantifies the population mismatch between the final
state $\rho_{\tau}$ and the equilibrium state $\rho_B$. This
population difference is expressed in terms of the dephasing map
in the instantaneous energy basis, $\Delta_{\tau}$, where for
$t\in [0, \tau]$,
\begin{equation}
\Delta_{t}[\rho]: =\sum_n \ket{n(t)}\bra{n(t)} \rho
\ket{n(t)}\bra{n(t)} \, .
\end{equation}
To prove Eq. (\ref{theo1}), we notice that the minimum in Eq.
(\ref{CoherMeas}) is achieved by
$\sigma=\Delta_{\tau}[\rho_{\tau}]$~\cite{coherRev} so that
\begin{equation}
C(\rho_\tau) = D(\rho_\tau||\Delta[\rho_\tau]) =
S(\Delta[\rho_\tau])-S(\rho_{\tau}) \, .
\end{equation}
Since $S(\rho_{\tau}) = S(\rho_0) \equiv S(\Delta_0[\rho_0])$, the
amount of coherence generated in the process, $C(\rho_\tau)$,
coincides with the production of diagonal entropy,  as introduced
by Polkovnikov \cite{polko11}. Then,
\begin{eqnarray}
\beta_i \langle w_{irr} \rangle &=& D(\rho_\tau||\rho_B) =-S(\rho_\tau) - \Tr{\rho_\tau \ln \rho_B} \nonumber \\
&=& -S(\rho_0) - \Tr{\rho_\tau \ln \rho_B} \nonumber \\
%&=& -S(\rho_0) + S(\Delta[\rho_\tau])-S(\Delta[\rho_\tau]) - \Tr{\rho_\tau \ln \rho_B}\\
&=& C(\rho_\tau) -S(\Delta[\rho_\tau]) - \Tr{\rho_\tau \ln \rho_B} \nonumber \\
%&=& C(\rho_\tau) -S(\Delta[\rho_\tau]) - \Tr{\Delta[\rho_\tau] \ln \rho_B} \nonumber \\
&=& C(\rho_\tau) + D(\Delta_{\tau}[\rho_\tau]||\rho_B) \, .
\label{unicaproof}
\end{eqnarray}
This equality has a simple physical interpretation: it simply tells us
that the amount of irreversible work performed on the system (and,
loosely speaking, the amount of irreversibility) can either go
into coherence generation, or it can produce non-equilibrium
populations (with respect to $\rho_B$).

All the steps of the proof above could be repeated for any
intermediate time $t \in [0, \tau]$, so that Eq. (\ref{theo1}) can
be generalized as follows:
\begin{equation}
\beta_i \langle w_{irr}(t) \rangle =  C(\rho_t) +
D(\Delta_{t}[\rho_t] ||\rho^{\beta}_{\lambda(t)}) \label{theogene}
\end{equation}
where  the irreversible work generated up to time $t$ is defined
as the average work performed up to $t$ minus the ``instantaneous
free energy difference'', $\langle w_{irr}(t) \rangle = \Tr{\rho_t
H[\lambda(t)] } - \Tr{\rho_0 H_i} - (F_{\beta_i,\lambda(t)}-
F_i)$. Eq. (\ref{theogene}) tells us that the decomposition of the
irreversible entropy production into a coherent and and incoherent
contributions can be performed at any time during the process.

{\bf Physical Examples} - To gain some insight into this
decomposition, we will now focus on studying it in two physically
motivated examples. First, we consider the example of a finite
time unitary process performed on a qubit. In particular, we focus
on the process experimentally realized in Ref. \cite{expe2}, for
which we calculated the irreversible work and the fraction of it
that is due to coherence generation. The results are reported in
Fig.~(\ref{scheme}).
\begin{figure}[b]
        \begin{center}
\includegraphics[width=0.8\columnwidth]{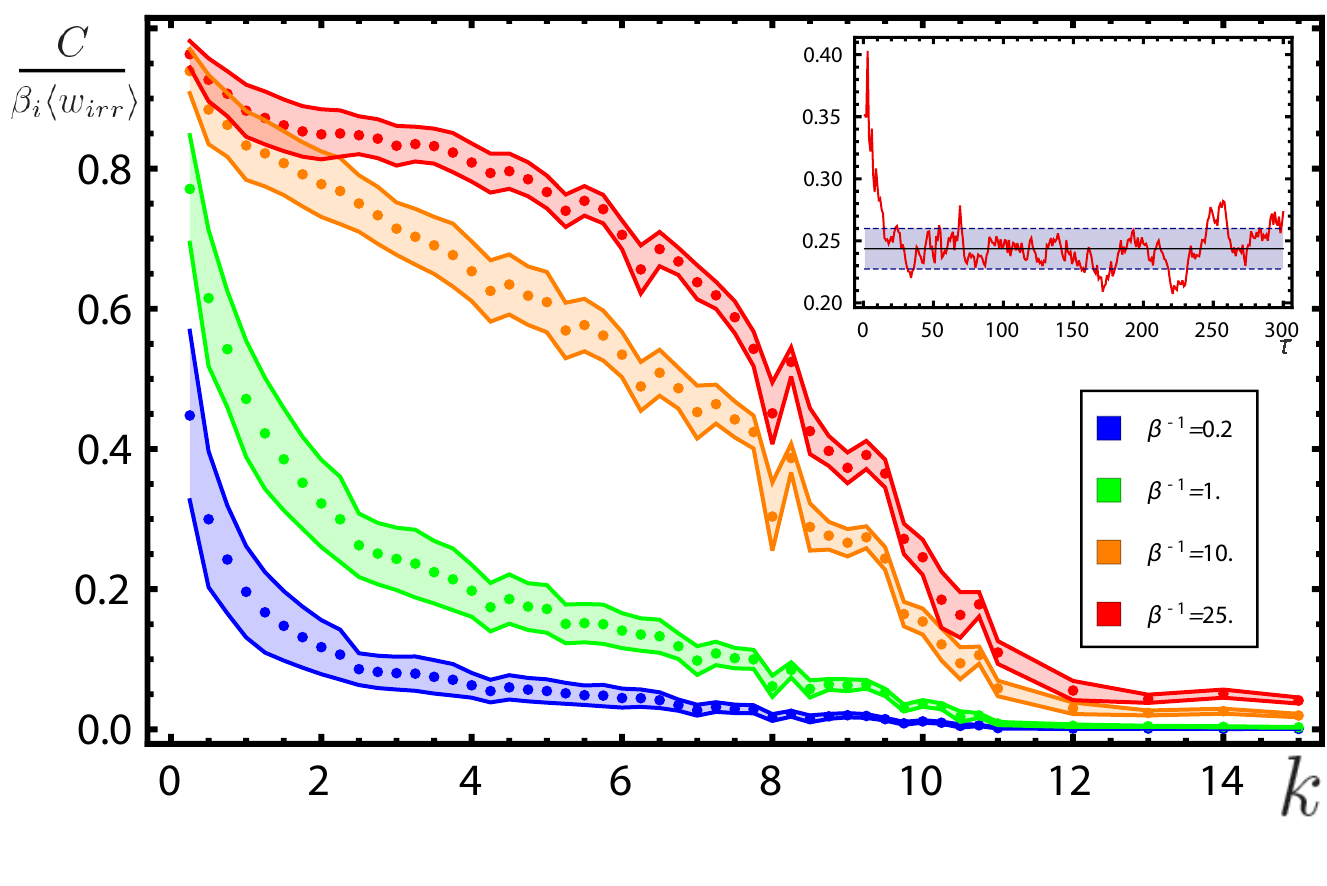}
         \includegraphics[width=0.8\columnwidth]{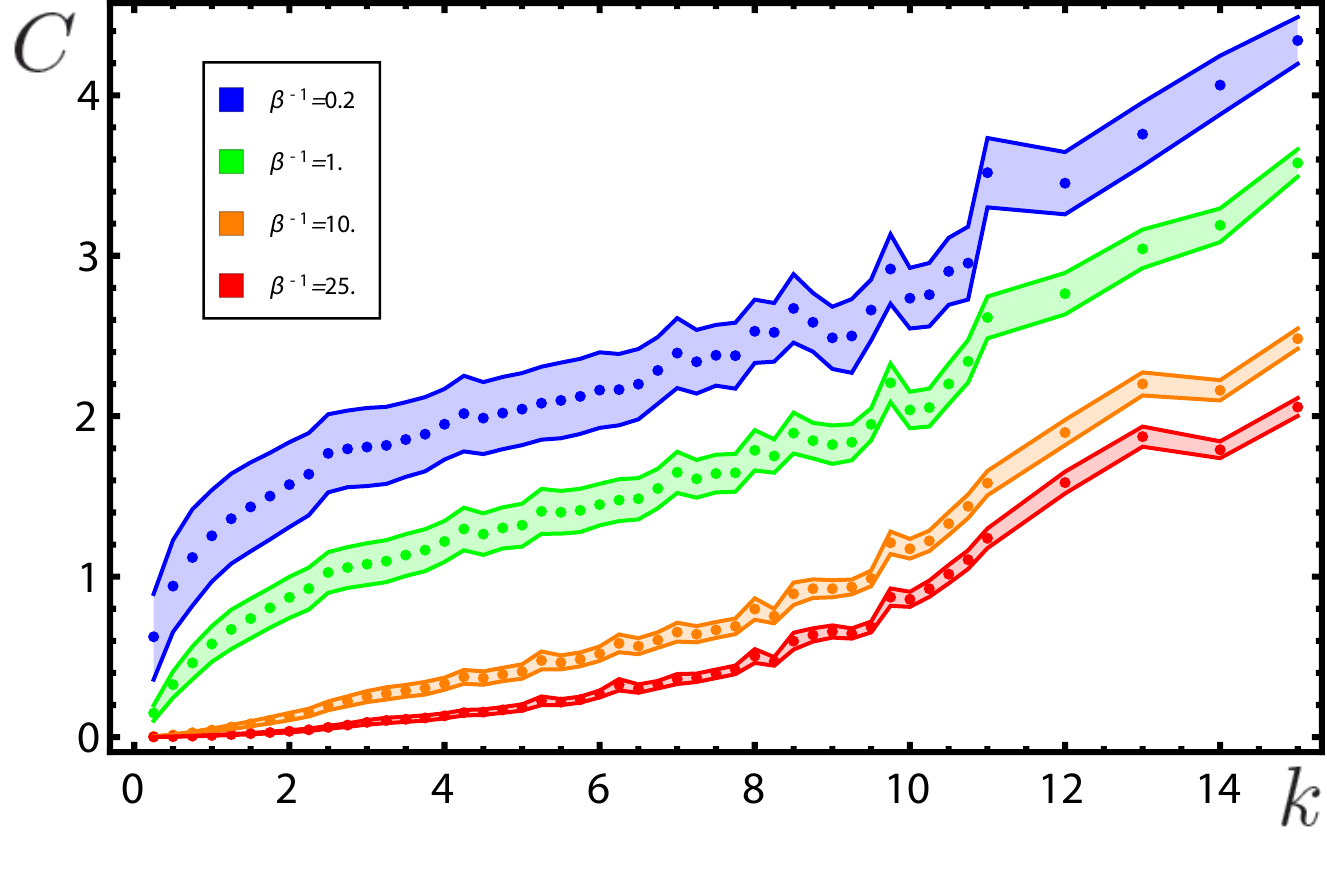}
        \caption{(Color online). {\it Upper panel} - Fraction of
the (irreversible) work that goes into the generation of quantum
coherence, shown as a function of the kick strength $k$, for
different values of the temperature $\beta^{-1}$ and with
$T=0.25$. The dots give the average saturation value, estimated by
a long-time average (performed over a time window $\Delta \tau
> 3000$ ), while the filled region has half width given by the
standard deviation of the fluctuations. The precise meaning of the
average and fluctuation estimation can be understood from the
inset, which displays the same ratio as a function of the number
of kicks $\tau$, for the specific case $k=9.5$ $\beta^{-1}=0.1$.
After an initial diffusion stage, a saturation regime is reached
where fluctuations with $\tau$ are observed (for both coherence
and average work) around the average values reported in the main
plot and with an amplitude given by the shaded region (which tends
to shrink with increasing the initial temperature). {\it Lower
panel} - Average saturation value of the coherence generated in
the process as a function of $k$, for $T=0.25$ and different
temperatures. The remaining irregular behavior of $C$ (and of
$\langle w_{irr} \rangle$) can be explained as in Ref.
\cite{irregular}. }
        \label{figkick}
        \end{center}
\end{figure}
In Fig. \ref{figkick}, we considered the more sophisticated
example of a quantum kicked rotor \cite{casatibook}, describing a
particle freely evolving on a circle, $\hat H_0= \frac{{\hat
p}^2}{2}$ with $\hat p=-i\partial_\theta$, and $\theta\in
[0,2\pi)$, brought far from its initial equilibrium state by a
series of kicks at intervals $T$, $H(t) = H_0 + V(\theta)
\sum_{n\in \Z}\delta(t-nT)$. In particular, we focus on
$V(\theta)= k \cos(\theta)$, leading to the  standard quantum map
\cite{Izrailev90}. The state after the $\tau^{th}$ kick will be
$\rho_\tau = U^\tau \rho_0 {U^\tau}^\dag$, with $\hat U= e^{-i
\hat H_0 T} e^{-i V(\hat\theta)}$, and with the initial
equilibrium state $\rho_0=\exp \{-\beta_i H_0\}/Z_0$.  Since the
Hamiltonian is periodic, the irreversible work coincides with the
average work itself, $\langle w_{irr}(\tau) \rangle \equiv \langle
w(\tau) \rangle = \Tr{H_0(\rho_\tau - \rho_0)} $. The kicks give
rise to a spreading of the particle distribution in momentum
space. After such a diffusive transient, and due to quantum
interference, a relaxation process takes place, giving rise to a
localization in momentum space over a width $\xi_p$
\cite{fishman},  and (on average over quick fluctuations) to a
saturation of the system's energy. Quantum coherence is essential
for such a localization process to occur, and, indeed, $C(\rho_t)$
is generated during the process. The lower panel of
Fig.~\ref{figkick} shows the amount of coherence generated as a
function of the kick strength, evaluated in the localized (long
time) regime, while the upper panel displays the ratio $C/ \langle
S_{irr}\rangle $, giving the fraction of irreversible work that is
used to produce coherence. More coherence is generated at lower
temperatures, but $C$ and $\beta \langle w_{irr} \rangle$ scale
differently with $\beta_i$ so that the ratio decreases when the
temperature is lowered. A simple estimate in terms of the
localization parameter gives $\langle w \rangle \sim \xi_p^2$,
while $C \sim \ln (1+ \beta\xi_p^2)$, with $\xi_p \sim k^2$ (for
large $k$'s).

{\bf Non-adiabaticity} - By looking at Fig. (\ref{scheme}), one
sees that the coherent contribution goes to zero with increasing
the process duration $\tau$. This is not an accident, and further
insights are gained by studying the energetic deviation from the
adiabatic process (rather than from the isothermal one). To this
end, we start by recalling the meaning of an ideal quantum
adiabatic process. Starting from the same initial equilibrium
state $\rho_0$, the adiabatic state is $\rho_A := U_A \rho_0
U_A^{\dag}$, where $U_A$ describes a process in which $\lambda_i$
changes into $\lambda_f$ with an infinitely slow rate, $U_A =
\lim_{\tau \rightarrow \infty} U_{\tau,0}[\lambda]$. Under the
premises of the quantum adiabatic theorem \cite{Messiah,verd}, the
system undergoes a transition-less evolution, and $ \rho_A =
\sum_{n} \rho_{A, nn} \ket{n(\tau)}\bra{n(\tau)}$, with $\rho_{A,
nn} = \rho_{nn}(0)$. Generally, this is not a thermal state.
However, in Ref. \cite{innerfriction}, it has been shown that, if
this is the case, then one can express the non-adiabatic part of
the work, also called {\it inner friction} \cite{kossloff}, in
terms of the non-adiabaticity parameter $\mathcal A :=
D(\rho_\tau||\rho_A)$, given by the relative entropy between
$\rho_\tau$ and $\rho_A$. This is an operationally useful quantity
in itself, which can serve as an entropic quantifier of the
deviation from the adiabatic evolution \cite{inprep}. It can be
relevant, e.g., to evaluate the performance of control methods
giving rise to shortcuts to adiabaticity
\cite{shortcutRev,Berry,shortcut1,shortcut2,gabriele}.

During the actual process, the state of the system is unable to
follow the instantaneous adiabatic state, as signaled by a non
zero $\mathcal A$, due to either the generation of coherence or
unwanted {\it diabatic} transitions. Indeed, as we did before for
the irreversible work, two contributions can be isolated in the
non-adiabaticity parameter $\mathcal A$ as well, giving
\begin{equation}
\mathcal A =  C(\rho_\tau) + D(\Delta_{\tau}[\rho_\tau]|| \rho_A)
\label{theo2}
\end{equation}
where, again, $ C(\rho_\tau)$ measures the amount of coherence
generated by the driving, while
$D(\Delta_{\tau}[\rho_\tau]||\rho_A)$ is the contribution to
non-adiabaticity given by population changes. The proof of this
relation is exactly the same as in Eq.~(\ref{unicaproof}), with
the only change of $\rho_A$ in place of $\rho_B$ in every step.

In fact, for slow enough (but yet non-adiabatic) processes, it is
the coherence that gives the leading contribution in Eq.
(\ref{theo2}). To show that this is indeed the case, let us
consider a slow enough evolution between $t=0$ and $t=\tau$, so
that $U_{\tau,0}[\lambda]$ can be approximated by its adiabatic
series expansion in powers of $1/\tau$ \cite{Messiah}. To second
order in $\tau^{-1}$, one has
\begin{equation}
\mathcal A = C(\rho_\tau) + {\cal O} \left ( \frac{1}{\tau^2}
\right ) \label{AcircaC} \, .
\end{equation}
The proof of this approximate relation goes as follows. Up to
second order in the adiabatic series, the transition probability
is given by \cite{Berry}
\begin{equation}
P_{m\rightarrow n}(\tau) \approx \delta_{m\,n} +
P^{(2)}_{m\rightarrow n}(\tau) \, .
\end{equation}
Using this approximation, we can express the diagonal entropy of
the actual final state, $S(\Delta[\rho_\tau])$, in the form
\begin{eqnarray*}
&& S(\Delta_{\tau}[\rho_\tau]) = - \sum_n \rho_{nn}(\tau) \ln \rho_{nn}(\tau)\\
&& \quad \approx - \sum_n \rho_{nn}(\tau) \ln\left( \rho_{nn}(0) + \sum_m \rho_{mm}(0) P^{(2)}_{m\rightarrow n}(\tau)\right)\\
&& \quad = - \sum_n \rho_{nn}(\tau) \ln \rho_{nn}(0) \\ && \quad
- \sum_n \rho_{nn}(\tau) \ln\left(  1 + \sum_m \frac{\rho_{mm}(0)}{\rho_{nn}(0)} P^{(2)}_{m\rightarrow n}(\tau)\right)\\
&& \quad \approx - \sum_n \rho_{nn}(\tau) \ln \rho_{nn}(0) -
\sum_n  \sum_m \rho_{mm}(0) P^{(2)}_{m\rightarrow n}(\tau)
\end{eqnarray*}
For the second order transition probabilities, the sum rule
$\sum_n P^{(2)}_{m\rightarrow n}(\tau) = 0$ holds, and therefore
\begin{eqnarray*}
S(\Delta_{\tau}[\rho_\tau]) &\approx& - \sum_n \rho_{nn}(\tau) \ln
\rho_{nn}(0) =  - \Tr{\rho_\tau \ln \rho_A}
\end{eqnarray*}
As a result,
\begin{eqnarray*}
    \mathcal A &=& -S(\rho_\tau) -  \Tr{\rho_\tau \ln \rho_A}\\
%     &=& -S(\rho_i) -  \Tr{\rho_\tau \ln \rho_A}\\
     &\approx& -S(\rho_i) + S(\Delta[\rho_\tau]) \equiv C(\rho_\tau) \, ,
\end{eqnarray*}
which completes the proof of Eq. (\ref{theo2}). This relation
implies that, as we start to deviate from the adiabatic evolution
(that would lead to the incoherent state $\rho_A$), the
non-adiabaticity parameter $\mathcal A$ becomes non-zero due to
coherence production, while diabatic transitions give sub-leading
contributions.

In the limit $\tau \rightarrow \infty$, $\mathcal A$ is zero (by
definition), and, thus, $C(\rho_{\tau})$ needs to decay as $\tau$
increases, as in Fig. (\ref{scheme}). The fact that $\beta_i
\langle w_{irr} \rangle$ persists in the long $\tau$ limit,
instead, is fully attributable to the second, population term in
Eq. (\ref{theo1}).

As a final note, for cyclic processes we have $\rho_A \equiv
\rho_B \equiv \rho_0$ and therefore the non-adiabaticity parameter
coincides with $\langle S_{irr} \rangle$.

{\bf Fluctuation relations} - In close analogy with the analysis
performed by Esposito and Van der Broeck \cite{storici}, we now
show that the separation of the irreversible entropy into a
contribution due to coherence generation and another one due to
the production of a population mismatch with respect to the
reference equilibrium state leads to three integral fluctuation
theorems. Adopting the (by now standard) two measurement
framework, one imagines to measure energy at $t=0$ obtaining the
result $\epsilon_n(0)$, and again, at time $t=\tau$, obtaining
$\epsilon_m(\tau)$ \cite{CampisiRev}. We can, then, define three
stochastic variables that take the values
\begin{eqnarray}
&& s_{nm} := \beta_i [(\epsilon_m(\tau) - \epsilon_n(0)) - (F_B -
F_i)] \, ,\\
&& p_{nm} := \ln \rho_{mm}(\tau) - \ln \rho_{B, mm} \, , \\
&& c_{nm} := s_{nm} - p_{nm} = \ln \rho_{nn}(0) - \ln
\rho_{mm}(\tau) \, .
\end{eqnarray}
If these are distributed according to the probability density
$$P(\alpha) = \sum_{n,m} \rho_{nn}(0) P_{n \rightarrow m}(\tau)
\, \delta (\alpha - \alpha_{nm} ) \, , \quad \mbox{for } \, \alpha
= s, p, c \, ,$$ then it is a matter of a simple algebra to show
that
\begin{equation}
\langle s \rangle \equiv \langle S_{irr} \rangle \, , \; \langle p
\rangle \equiv D(\Delta_{\tau}[\rho_{\tau}] || \rho_B) \, , \;
\langle c \rangle \equiv C(\rho_{\tau}) \, ,
\end{equation}
where the average values are calculated as $\langle \alpha \rangle
= \int \alpha P(\alpha) d \alpha $. Indeed, $s$ is just the usual
stochastic variable giving irreversible work divided by the
initial temperature, $p$ describes the final population mismatch,
while $c$ is the difference between the previous two. (Notice that
$n$ is a dummy index in $p_{nm}$ and that, accordingly, the
summation over $n$ in $P(p)$ is un-effective).

These three variables, whose averages combine as $\langle s
\rangle = \langle p \rangle + \langle c \rangle$, satisfy the
fluctuation relations
\begin{equation}
\langle e^{-s} \rangle = \langle e^{-c} \rangle = \langle e^{-p}
\rangle = 1 \, .\end{equation} The division of the irreversible
entropy production in the two `basic' contributions due to
coherence and to population imbalance generation is, thus, in some
sense, `natural' in this stochastic framework, as the two parts
satisfy independent fluctuation theorems as well. In particular,
since $C(\rho_{\tau})$ coincides with the increase in
Polkovnikov's diagonal entropy, the relation $\langle e^{-c}
\rangle =1$ provides a fluctuation theorem for it which is quite
remarkable and suggests an interesting, uniquely quantum aspect to
certain transformations in non equilibrium thermodynamics.

{\bf Conclusions} - Let us summarize our results. We have shown,
using the concept of relative entropy of coherence, that a
uniquely coherent contribution to the energetics of a driven
quantum system can be identified. Given the current interest and
development of thermodynamics of non equilibrium quantum systems,
this is an important result and can be used to analyze in detail
the role of coherence in thermodynamic transformations. We have
illustrated our result with two physical examples. Furthermore, we
have shown that this coherence term appears also in the
quantification of non-adiabaticity and that, up to second order in
the adiabatic expansion, the coherence is actually a measure of
the deviation from adiabatic evolution. Furthermore, in this
framework, coherence generation coincides with the increase in
diagonal entropy, and, thus, our analysis clarifies the role of
the latter in quantum thermodynamics. Finally we have shown that
the coherence itself obeys a fluctuation theorem.

\noindent {\it Acknowledgements} G.F. and F.P. acknowledge support
from the Collaborative Project QuProCS (Grant Agreement 641277). J.G. would like to thank A. Silva for useful discussions.

\end{document}